# Beyond Node Degree: Evaluating AS Topology Models

Hamed Haddadi[*]
University College London

Damien Fay
University of Cambridge

Almerima Jamakovic
Delft University of Technology

Olaf Maennel[†]
Deutsche Telekom Laboratories

Andrew W. Moore
University of Cambridge

Richard Mortier[‡]
Vipadia Ltd

Miguel Rio
University College London

Steve Uhlig
Delft University of Technology

## ABSTRACT

Many models have been proposed to generate Internet Autonomous System (AS) topologies, most of which make structural assumptions about the AS graph. In this paper we compare AS topology generation models with several observed AS topologies. In contrast to most previous works, we avoid making assumptions about which topological properties are important to characterize the AS topology. Our analysis shows that, although matching degree-based properties, the existing AS topology generation models fail to capture the complexity of the local interconnection structure between ASs. Furthermore, we use BGP data from multiple vantage points to show that additional measurement locations significantly affect local structure properties, such as clustering and node centrality. Degree-based properties, however, are not notably affected by additional measurements locations. These observations are particularly valid in the core. The shortcomings of AS topology generation models stems from an underestimation of the complexity of the connectivity in the core caused by inappropriate use of BGP data.

## Categories and Subject Descriptors

C.2.1 [**Network Architecture and Design**]: Network topology; I.6.4 [**Simulation and Modeling**]: Model Validation and Analysis

[*]This work was done while the author was visiting the Computer Laboratory, University of Cambridge.
[†]This work was done while the author was at School of Mathematical Sciences, University of Adelaide.
[‡]This work was done while the author was at Microsoft Research Cambridge.

## 1. INTRODUCTION

For many years researchers have modeled the Internet's Autonomous System (AS) topology[1] using graphs obtained via various measurement techniques, e.g. BGP routing tables [16, 28] and traceroute maps [18]. The AS topology is an abstraction of the Internet which is commonly used to analyze its characteristics and simulate the performance and scalability of new protocols and applications. Simulation methods require that AS topology generation models be able to provide topologies whose properties are as close as possible to those of the observed AS topology.

In this paper we evaluate existing AS topology generation models by comparing them with several available datasets, representing observed AS topologies of the Internet. Figure 1 illustrates the relationship between the Internet topology, its measurement instances, and AS topology generation models.

A key principle underlying our work is to be agnostic about the topological properties of the Internet. The main reason for our agnosticism lies in the dynamic behavior of the Internet topology. In addition, observations of the AS topology suffer from two problems. On the one hand, common set of observation points have only limited visibility of the topology [26]. On the other hand, each observation technique suffers from measure-ment artifacts. This results in problems for BGP-based as well as traceroute-based observations of the Internet topology. For example, traceroute can report hops that do not map to a unique AS number [22]. As a result, AS topology models make use of simplifying assumptions about the actual topology [5, 19, 37]. One widely held assumption, based on biased observations, is that the AS topology has a hierarchical structure [30] and its node-degree distribution obey a power-law [12].

---
[1]Note that the AS topology neither represents the dataplane topology nor directly corresponds to the Internet router-level topology. Many organizations are permanently connected to their providers, sharing an AS number [29]. Alternately, a single organization may use many AS numbers for controlling routing.

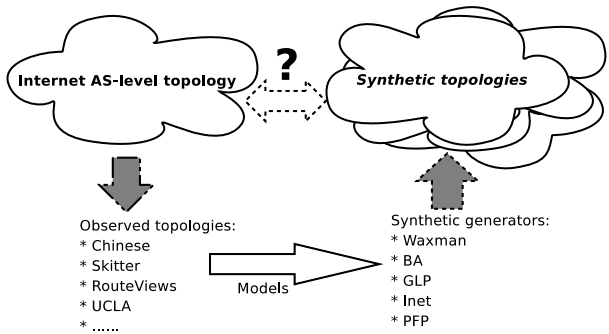

Figure 1: Internet topology generation

Believing that at present it is impossible to know better, we accept the fact that the AS topology observations suffer from biases and thus reveal different partial truths about the properties of the Internet. However, comparison of different observed AS topologies with different levels of incompleteness, and topologies generated from different models, allows us to learn from the limitations of particular assumptions about the Internet's AS topology. Then, the direction of these biases and limitations may gives us insight into the actual properties of the AS topology.

To evaluate AS topology generation models, we rely on a wide set of commonly used topological metrics. We do not claim that the set of considered metrics captures all important aspects of the AS topology. However, using such an extensive set of topological metrics allows us to observe differences in the so far revealed topological properties of observed and sythetic AS topologies. Futhermore, we rely on statistical measures to compare distributions of some metrics, allowing us to measure more objectively the similarity of two topologies.

In this paper we show that the existing topology generators capture the node degree distributions quite well, but fail to account either for the complex local interconnection structure between ASes, or the highly meshed structure of the core AS topology. Such shortcomings can affect the performance of protocols and applications when simulated using synthetic topologies. We also show that, using additional BGP peering vantage points for collecting connectivity information, does greatly affect important characteristics, such as power-laws and measures of centrality, while having little effect on basic degree-related properties. These observations suggest that for understanding the nature of the Internet topology one should use rich(er) datasets, which capture a large portion of existing peering links. Moreover, they show that the significance of preferential-attachment has waned while peering links, underestimated in the past, are now far more important.

The rest of this paper is structured as follows. In Section 2 we contrast our work with the related work. We then introduce the existing AS topology generation models and describe their underlying assumptions in Section 3, and present a set of observed AS topologies collected using different methodologies from various locations in the world in Section 4. Subsequently, in Section 5 we describe the metrics used for topology characterization and discuss the statistical measures of similarity in Section 6. In Section 7, we present the results of our comparison analysis. Due to the discovery that synthetic and observed topologies record biases related to the nature of the data collection processes, we conduct an extensive analysis of the impact of increasing the number of BGP peering vantage points on our topology dataset, collected from a large number of measurement locations. This study is presented in Section 8. Finally, in Section 9 we conclude and discusses potential improvements in the field of AS topology modeling.

## 2. RELATED WORK

Zegura *et al.* [35] analyse topologies of 100 nodes generated using pure-random, Waxman [32], exponential and several locality based models of topology such as Transit-Stub. They use metrics such as average node degree, network diameter, number of paths between nodes. They find that pure random topologies represent expected properties such as locality very poorly and so we exclude them from our comparison. They suggest that the Transit-Stub method should be used due both to its efficiency and the realistic average node degree its topologies achieve.

Faloutsos *et al.* [12] state that three specific properties of the Internet AS topology are well described by power laws: rank exponent, out-degree exponent and eigen-exponent (graph eigenvalues). This work parallelled development of many models incorporating power laws, such as the Barabási and Albert [3] model, based on incremental growth by addition of new nodes and preferential attachment of new nodes to existing well-connected nodes.

Later, Bu and Towsley [5] used the empirical complementary distribution (ECD) rather than standard histograms to generate new nodes. They showed the variability in graphs from different generators using the same heuristics using characteristic path length and clustering coefficients.

Tangmunarunkit *et al.* [31] provide a first comparison of the underlying characteristics of degree-based models against structural models. A major conclusion is that the the simplest form of degree-based model performs better than random or structural models at representing the studied parameters. They compare three categories of model generators: Waxman, Tiers [10] and the Transit-Stub structural model, against the simplest degree based generator, the power-law random graph (PLRG) [1]. They define and use three metrics: expansion, resilience and distortion. They find that the PLRG matches these metrics better than the random or structural models. Based on these metrics they conclude that stricter hierarchy is present in the measured networks than in degree-based generators. However, they leave many questions unanswered about the accuracy of degree-based generators and the choice of metrics.

Zhou and Mondragon [37] propose models based on several mathematical features, such as rich-club, interactive growth and betweenness centrality. They use AS data from the CAIDA Skitter project to examine the Joint Degree Distribution (JDD) and rich-club connectivity. They show that for these data, rich-club connectivity and the JDD are closely linked for a network with a given degree distribution.

In this paper, we consider more recent degree-based generators using a larger set of graph-theory derived metrics to give better insight into correct understanding of the AS topology. We compare in detail against a range of different Internet AS topologies at national and international level obtained from traceroute and BGP data. When choosing our metrics, we considered both metrics used by the topology generator designers and those used more widely in graph theory. A particular point to note is that we chose not to use the three metrics of Tangmunarunkit *et al.* for two reasons. First, computation of both resilience and distortion are NP-complete, requiring use of heuristics. In contrast, all our metrics are straightforward to compute directly. Second, although accurate reproduction of degree-based metrics is well-supported by current topology generators, our hypothesis was that local interconnectivity was poorly supported, and so we chose to use several metrics that focus on exactly this, e.g., assortativity, clustering, and centrality.

## 3. AS TOPOLOGY MODELS

There are many models available that claim to describe the Internet AS topology. Several of these models are embodied in tools for generating simulated topologies [15]. In this section we describe the particular models whose output we compare in this paper. The first are produced from the Waxman model [32], derived from the Erdös-Rényi random graphs [11], where the probability of two nodes being connected is proportional to the Euclidean distance between them. The second come from the Barabasi and Albert [3] model, following measurements of various power laws in degree distributions and rank exponents by Faloutsos *et al.* [12]. These incorporate common beliefs about preferential attachment and incremental growth. The third are from the Generalized Linear Preference model [5] which additionally model clustering coefficients. Finally, Inet [33] and PFP [37] focus on alternative AS topology characteristics: the meshed core and preferential attachment respectively. Each model focused only on particular metrics and parameters, and only compared their output with selected AS topology observations.

### 3.1 Waxman

The Waxman model of random graphs is based on a probability model for interconnecting nodes of the topology given by: $P(u,v) = \alpha \ e^{-d/(\beta L)}$, where $0 < \alpha, \beta \leq 1$, $d$ is the Euclidean distance between two nodes $u$ and $v$, and $L$ is the network diameter (largest distance between two nodes). We use the BRITE [23] implementation of this model in this paper, which facilitates rewiring using iterative assignment of edges to ensure that there are no disconnected components in the generated topology.

### 3.2 BA

The BA model was inspired by the idea of preferentially attaching new nodes to existing well-connected nodes, leading to the incremental growth of nodes and the links between them. When a node $i$ joins the network, the probability that it connects to a node $j$ already in the network is given by: $P(i,j) = \dfrac{d_j}{\sum_{k \in V} d_k}$ where $d_j$ is the degree of node $j$, $V$ is the set of nodes that have joined the network and $\sum_{k \in V} d_k$ is the sum of degrees of all nodes that previously joined the network [23].

### 3.3 GLP

Our third model is the Generalized Linear Preference model (GLP) [5]. This focuses on matching characteristic path length and clustering coefficients. It uses a probabilistic method for adding nodes and links recursively while preserving selected power law properties.

### 3.4 Inet

Inet [33] produces random networks using a preferential linear weight for the connection probability of nodes after modelling the core of the generated topology as a full mesh network. Inet sets the minimum number of nodes at 3037, the number of ASs on the Internet at the time of its development. It similarly sets the fraction of degree 1 nodes to 0.3, based on measurements from Routeviews[2] and NLANR[3] BGP table data in 2002.

## 3.5 PFP

In the Positive Feedback Preference (PFP) model, the AS topology of the Internet is considered to grow by interactive, probabilistic addition of new nodes and links. It uses a nonlinear preferential attachment probability when choosing older nodes for the interactive growth of the network, by inserting edges between existing nodes as well as the newly added ones.

## 4. AS TOPOLOGY OBSERVATIONS

The Internet AS topology can be inferred from various sources of data such as BGP or traceroute [21] at the network (IP) layer. Using BGP routing data alone suffers from incompleteness, no matter how many vantage points are used to collect observations. In particular, even if BGP updates are collected from multiple vantage points and combined, many peering and sibling relationships are not observed [13]. Conversely, traceroute data misses alternative paths since routers may have multiple interfaces which are not easily identified, and multi-hop paths may also be hidden by traffic tunnelled via Multi-Protocol Label Switching (MPLS). Combining these data sources still does not solve all problems since mapping traceroute data to AS numbers is not always accurate [22]. In this paper we attempt to avoid these problems by comparing against many measurement-derived datasets giving a diverse spatial and temporal comparison across different continents and years of measurement.

### 4.1 Chinese

The first dataset is a traceroute measurement of the Chinese AS Topology collected from servers within China in May 2005. It reports 84 ASs, representing a small subgraph of the Internet. Zhou et al. [38] maintain that the Chinese AS graph presents all the major topology characteristics of the global AS graph. The presence of this dataset enables us to compare the AS topology models at smaller scales. Further, this dataset is believed to be nearly complete, i.e., it contains very little measurement bias and accurately represents the true AS topology for that region of the Internet. Thus, although it is rather small, we have included it as a valuable comparison point in our studies.

### 4.2 Skitter

The second dataset comes from the CAIDA Skitter project[4]. CAIDA computes the adjacency matrix of the AS topology from the daily Skitter measurements. These are obtained by running traceroutes over a large range of IP addresses and mapping the prefixes to AS numbers using RouteViews BGP data. Since this data reports paths actually taken by packets, rather than path information propagated via BGP, it more directly represents the IP topology than the BGP data alone. For our study, we used the graphs for March 2004 as used by Mahadevan et al. [20], which reports 9,204 unique ASs.

### 4.3 RouteViews

The third dataset we use is derived from the Route-Views BGP data. This is collected both as static snapshots of the BGP routing tables and dynamic BGP data in the form of BGP update and withdrawal messages. We use the topologies provided by Mahadevan et al. [20] from two types of BGP data from March 2004: one from the static BGP tables and one from the BGP updates. In both cases, they filter AS-sets and private ASs and merge the 31 daily graphs into one. This dataset reports 17,446 unique ASs across 43 vantage points in the Internet.

### 4.4 UCLA

The fourth dataset comes from the Internet topology collection[5] maintained by Oliviera et al. [27]. These topologies are updated daily using the data sources such as BGP routing tables and updates from RouteViews, RIPE[6], Abilene[7] and LookingGlass servers. Each node and link is annotated with the times it was first and last observed. We use a snapshot of this dataset from November 2007 computed using a time window on the last-seen timestamps to discard ASs which have not been seen for more than 6 months. The resulting dataset reports 28,899 unique ASs.

## 5. TOPOLOGY CHARACTERIZATION

Over the past several years a veriety of topological metrics has been proposed to quantitatively characterize topological properties of networks. In this section we present a large set of topological metrics that will be used to measure a *distance* in graph space, i.e. how distant two graphs are topologically from each other. The topological metrics are computed for the synthetic and the measured AS topologies. Taken individually, these metrics concentrate on differing topological aspects but when considered together they reveal the shortcomings of topology models to faithfully capture the topological properties of observed AS topologies. AS topologies are modeled as graphs $G = (\mathcal{N}, \mathcal{L})$ with a collection of nodes $\mathcal{N}$ and a collection of links $\mathcal{L}$ that connect a pair of nodes. The number of nodes and links in a graph is then respectively equal to $N = |\mathcal{N}|$ and $M = |\mathcal{L}|$.

---

[2] http://www.routeviews.org/
[3] http://www.nlanr.net/
[4] http://www.caida.org/tools/measurement/Skitter/
[5] http://irl.cs.ucla.edu/topology/
[6] http://www.ripe.net/db/irr.html
[7] http://abilene.internet2.edu/

## 5.1 Degree

The degree $k$ of a node is the number of links adjacent to it. The *average node degree* $\bar{k}$ is defined as $\bar{k} = 2M/N$. The *node degree distribution* $P(k)$ is the probability that a randomly selected node has a given degree $k$. The node degree distribution is defined as $P(k) = n(k)/N$ where $n(k)$ is the number of nodes of degree $k$. The *joint degree distribution* (JDD) $P(k, k')$ is the probability that a randomly selected pair of connected nodes have degrees $k$ and $k'$. A summary measure of the joint degree distribution is the average neighbor degree of nodes with a given degree $k$, and is defined as follows $k_{nn}(k) = \sum_{k'=1}^{k_{max}} k' P(k'|k)$. The maximum possible $k_{nn}(k)$ value is $N - 1$ for a maximally connected network, i.e. a complete graph. Hence, we represent JDD by the normalized value $k_{nn}(k)/(N-1)$ [20] and refer to it as *average neighbor connectivity*.

## 5.2 Assortativity

Assortativity is a measure of the likelihood of connection of nodes of similar degrees [25]. This is usually expressed by means of the *assortativity coefficient* $r$: assortative networks have $r > 0$ (disassortative have $r < 0$ resp.) and tend to have nodes that are connected to nodes with similar (dissimilar resp.) degree.

## 5.3 Clustering

Given node $i$ with $k_i$ links, these links could be involved in at most $k_i(k_i - 1)/2$ triangles (e.g. nodes $a \to b \to c \to a$ form a triangle). The greater the number of triangles the greater the clustering of this node. The clustering coefficient, $\gamma(G)$, is defined as the average number of 3-cycles (i.e., triangles) divided by the total number of possible 3-cycles:

$$\gamma(G) = 1/N \sum_i \frac{T_i}{k_i(k_i - 1)/2}, k_i \geq 2$$

where $T_i$ is the number of 3-cycles for node $i$, $k_i$ is the degree of node $i$. We use the distribution of *clustering coefficients* $C(k)$, which in fact is the distribution of the terms $\frac{T_i}{k_i(k_i-1)/2}$ in the overall summation. This definition of the clustering coefficient gives the same weight to each triangle in the network, irrespective of the distribution of the node degrees.

## 5.4 Rich-Club

The *rich-club coefficient* $\phi(\rho)$ is the ratio of the number of links in the component induced by the $\rho$ largest-degree nodes to the maximum possible links $\rho(\rho - 1)/2$ where $\rho = 1...n$ are the first $\rho$ nodes ordered by their non-increasing degrees in a network of size $n$ nodes [8].

## 5.5 Shortest path

The shortest path length distribution $P(h)$, as commonly computed using Dijsktra's algorithm, is the probability distribution of two nodes being at minimum distance $h$ hops from each other. From the shortest path length distribution, the average node distance in a connected network is derived as $\bar{h} = \sum_{h=1}^{h_{max}} hP(h)$, where $h_{max}$ is the longest among the shortest paths between any pair of nodes. $h_{max}$ is also referred to as the diameter of a network.

## 5.6 Centrality

Betweenness centrality is a measure of the number of shortest paths passing through a node or link, a centrality measure of a node or link within a network. The *node betweenness* for a node $v$ is $B(v) = \sum_{s \neq v \neq t \in \mathcal{N}} \frac{\sigma_{st}(v)}{\sigma_{st}}$ where $\sigma_{st}$ is the number of shortest paths from $s$ to $t$ and $\sigma_{st}(v)$ is the number of shortest paths from $s$ to $t$ that pass through a node $v$ [17]. The average node betweenness is the average value of the node betweenness over all nodes.

Closeness is a another measure of the centrality of a node within a network. The closeness of a node is the reciprocal of the sum of shortest paths from this node to all other reachable nodes in a graph.

## 5.7 Coreness

The $l$-core of a network (also known as the $k$-core) is the maximal component in which each node has at least degree $l$. In other words, the $l$-core is defined as the component of a network obtained by recursively removing all nodes of degree less than $l$. A node has coreness $l$ if it belongs to the $l$-core but not to the $(l + 1)$-core. Hence, the $l$-core layer is the collection of all nodes having coreness $l$. The core of a network is the $l$-core such that the $(l + 1)$-core is empty [4].

## 5.8 Clique

A clique in a network is a set of pairwise adjacent nodes, i.e. a component which is a complete graph. The *top clique size*, also known as the graph clique number, is the number of nodes in the largest clique in a network [34].

## 5.9 Spectrum

Recently, it has been observed that eigenvalues are closely related to almost all critical network characteristics [7]. For example, Tangmunarunkit *et al.* [31] classified network resilience as a measure of network robustness subject to link failures, resulting in a minimum balanced cut size of a network. Spectral graph theory enables studying this problem of network partitioning by using graph's eigenvalues [7]. In this paper we focus on *graph's spectrum*, i.e. the set of eigenvalues of the adjacency, the Laplacian or any other characteristic matrix of a graph. In the graph theory literature, one usually considers the adjacency or the Laplacian matrix [24, 9], both which employ different normalizations and therefore lead to different spectra. Here we focus on the spectrum of the *normalized Laplacian matrix* [7] where all eigenvalues lie between 0 and 2, allowing easy comparison of networks of different sizes. The normalized graph's spectrum has been successfully used for tuning the topology generators [14].

## 6. MEASURES OF SIMILARITY

To compare the distributions of various metrics we use the following statistics to determine how close two distributions are to each other. We perform the calculations for each synthetic topology instance separately and compare them to observed topologies of the same size. Note that distances are relative to the metric and the topology size, so that distances of one metric for a particular sized topology cannot be compared either to distances of another metric for the same sized topology, or to distances for the same metric for different sized topologies.

### 6.1 Kolmogorov-Smirnov (KS) distance

Given samples of two random variables, $X_1$ and $X_2$, the KS distance is the maximum empirical distribution difference defined as:

$$D_{max} = \sup |F_{n_1}(x) - F_{n_2}(x)|$$

where $F_{n_i}(x)$ is the empirical distribution of $X_i (i = 1, 2)$:

$$F_{n_i}(x) = \frac{1}{n_i} \sum_{j=1}^{n_i} I_{X_j \leq x} \text{ for } i\text{=1,2}$$

where $n_1$ and $n_2$ are the number of samples from $X_1$ and $X_2$ and $I_{X_j}$ is the indicator function.

The closely related 2-sample KS test tests the null hypothesis that $X_1$ and $X_2$ share a (true) *common* distribution based on the KS distance ($D_{max}$). However, it is misleading to use this test to indicate whether two distributions are *similar*, as it is highly sensitive to large sample sizes, and also as the particular $x_1$ and $x_2$ compared here are not strictly independent variables since, e.g., nodes with high degrees tend to occur together. Instead $D_{max}$ alone is used in this paper to indicate the *relative* closeness of distributions.

### 6.2 Kullback-Leibler divergence

The Kullback-Leibler (KL) divergence is also proposed as a suitable metric[8] for comparing network distributions. The KL divergence between two discrete random variables $X_1$ and $X_2$ is defined as:

$$D_{KL}(X_1, X_2) = \sum_i P(X_1 = X_i) log \frac{P(X_1 = X_i)}{P(X_2 = X_i)}$$

---
[8]The KL divergence is not strictly a metric as $D_{KL}(X_1, X_2) \neq D_{KL}(X_2, X_1)$

where $P(x)$ is the probability of $x$.

The KL divergence takes into account the difference between the distributions at all points rather than simply at the maximum point. In this paper, Gaussian kernel density estimation using fixed bins centered around data in the observed data set were found to perform well, although other methods do exist.

## 7. RESULTS AND DISCUSSION

Most past comparisons of topology generators have been limited to the average node degree, the node degree distribution and the joint degree distribution. The rationale for choosing these metrics is that if those properties are closely reproduced, then the value of other metrics will also be closely reproduced [19].

In this section we show that current topology generators are able to match first and second order properties well, i.e., average node degree and node degree distribution, but fail to match many other topological metrics. We also discuss the importance of various metrics in our analysis.

### 7.1 Methodology

For each generator we specify the required number of nodes and generate 10 topologies of that size in order to provide confidence intervals for the metrics. We then compute the values of the metrics introduced in Section 5 for the generated and observed AS topologies. It is important to note that all topologies studied in this paper are undirected, preventing us from considering peering policies and provider-customer relationships. This limitation is forced upon us by the design of the generators as they do not take such policies into account.

Each topology generator uses several parameters, all of which could be tuned to best fit a particular size topology, e.g., the Skitter dataset. However, there are two problems with attempting this tuning. First, doing so requires selection of an appropriate goodness-of-fit measure of which there are many, e.g., as noted in Section 5. Second, in any case tuning parameters to a particular dataset is of questionable merit since, as we argue in Section 1, each dataset is only a sample of reality with multiple biases and inaccuracies. Typically topology generator parameters are tuned so as to match the number of links in the synthetic and measured networks, for a given number of nodes. However we discovered this method to be inefficient as generating graphs with equal numbers of links from a random model and a power-law model gives completely different outputs. For space reasons we dealt with this particular issue elsewhere [14] and in this paper we simply use the default values embedded within each generator.

### 7.2 Topological metrics

Table 1: Comparison of AS level dataset with synthetic topologies.

| Topology | Links | Avg. deg. | Max. degree | Top clique size | Max. betweenness | Max. coreness | Assort. coef. | Clust. coef. | Max. closeness |
|---|---|---|---|---|---|---|---|---|---|
| *Chinese* | *211* | *5.02* | *38* | *2* | *1,324* | *5* | *-0.32* | *0.188* | *<0.01* |
| Waxman | 252 | 6 | 18 | **2** | 404 | 4 | 0.039 | 0.117 | 0.506 |
| BA | 165 | 3.93 | 19 | 3 | **1,096** | 2 | -0.096 | 0.073 | 0.515 |
| GLP | 151 | 3.6 | 44 | 3 | 2,391 | **5** | -0.257 | **0.119** | 0.643 |
| PFP | **250** | **5.95** | **37** | 10 | 849 | 9 | **-0.38** | 0.309 | 0.638 |
| *Skitter* | *28,959* | *6.3* | *2,070* | *16* | *10,210,533* | *28* | *-0.23* | *0.026* | *<0.01* |
| Waxman | **27,612** | **6** | 33 | 0 | 474,673 | 4 | 0.205 | 0.002 | **0.264** |
| BA | 18,405 | 4 | 190 | 0 | 5,918,226 | 2 | -0.05 | 0.001 | 0.315 |
| GLP | 16,744 | 3.64 | **2,411** | 2 | 34,853,544 | 5 | -0.089 | 0.003 | 0.496 |
| INET | 18,504 | 4.02 | 1,683 | 3 | 15,037,631 | 7 | -0.195 | 0.004 | 0.514 |
| PFP | 27,611 | **6** | 3,000 | **16** | **13,355,194** | **24** | **-0.244** | **0.012** | 0.588 |
| *RouteViews* | *40,805* | *4.7* | *2,498* | *9* | *30,171,051* | *28* | *-0.19* | *0.02* | *<0.01* |
| Waxman | 52,336 | 6 | 35 | 0 | 1,185,687 | 4 | 0.205 | 0.001 | **0.25** |
| BA | 34,889 | 4 | 392 | 3 | 33,178,669 | 2 | -0.04 | 0.001 | 0.33 |
| GLP | 31,391 | 3.6 | 4,226 | 4 | 127,547,256 | 6 | -0.08 | 0.002 | 0.48 |
| INET | **43,343** | **4.97** | **2,828** | **6** | **31,267,607** | 14 | -0.258 | 0.006 | 0.522 |
| PFP | 52,338 | 6 | 4,593 | 23 | 39,037,735 | **30** | **-0.252** | **0.009** | 0.564 |
| *UCLA* | *116,275* | *8.05* | *4,393* | *10* | *76,882,795* | *73* | *-0.165* | *0.05* | *0.32* |
| Waxman | 86,697 | 6 | 40 | 0 | 3,384,114 | 4 | 0.213 | <0.001 | 0.246 |
| BA | 57,795 | 4 | 347 | 0 | 52,023,288 | 2 | -003 | <0.001 | **0.3** |
| GLP | 52,456 | 3.63 | 7391 | 2 | 371,651,147 | 6 | -0.08 | <0.001 | 0.486 |
| INET | **91,052** | **6.3** | **6,537** | **12** | **88,052,316** | 38 | -0.3 | **0.01** | 0.55 |
| PFP | 86,696 | 6 | 8076 | 26 | 123,490,676 | **40** | **-0.218** | **0.01** | 0.57 |

In this section we discuss the results for each metric separately and analyze the reasons for differences between the observed and the generated topologies.

Table 1 displays the values of various metrics (columns) computed for different topologies (rows). Blocks of rows correspond to a single observed topology and the generated topologies with the same number of nodes as the observed topology. Bold numbers represent nearest match of a metric value to that for the relevant observed topology. Rows in each block are ordered with the observed topology first, followed by the generated topologies from oldest to newest generator. For synthetic topologies, the value of the metrics is averaged over the 10 generated instances. Note that Inet requires the number of nodes to be greater than 3037 and hence cannot be compared to the Chinese topology.

We observe a small but measurable improvement from older to newer generators in how well they match some metrics such as maximum degree, maximum coreness, and assortativity coefficient. This suggests that topology generators have been successively improved to better match some properties of the observed topologies. However, the number of links in the generated topologies may differ considerably from the observed topology due to the assumptions made by the generators. The Waxman and BA generators fail to capture the maximum degree, the top clique size, maximum betweenness and coreness. Those two generators are too simplistic in the assumptions they make about the connectivity of the graphs to generate realistic AS topologies. Waxman relies on a random graph model which cannot capture the clique between tier-1 ASs nor the heavy tail of the node degree distribution. BA tries to reproduce the power-law node degrees with its preferential attachment model but fails to reach the maximum node degree, as it only adds edges between new nodes and not between existing ones. Hence, neither of these two models is able to create the highly-connected core of tier-1 ASs. PFP and Inet manage to come closer to the values of the metrics of the observed topologies. For Inet this is because it assumes that 30% of the nodes are fully meshed (at the core), whereas for PFP its rich-club connectivity model allows to add edges between existing nodes.

### 7.2.1 Node degree distribution

In Figure 2 we show the CCDF of the node degree for all topologies on a log-log scale. We observe that the Chinese topology does not exhibit power law scaling due to its limited size, whereas all the larger AS topologies do exhibit power-law scaling of node degrees. The Waxman generator completely fails to capture this behavior as it is based on a random graph model, but recent topology generators do capture this power law behavior of the node degrees quite well. In the case of the RouteViews and UCLA datasets, Inet and PFP outperform other topology generators. Note that, contrary to RouteViews where the degree distribution displays strict power law scaling, the UCLA dataset has a slightly concave shape. In summary, more recent generation models reproduce node degree distribution well, as expected since most focus has been on this metric.

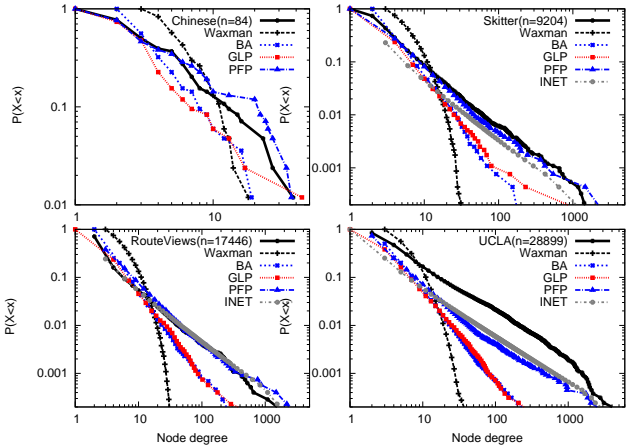

**Figure 2: Comparison of node degree CCDFs.**

### 7.2.2  Average neighbor connectivity

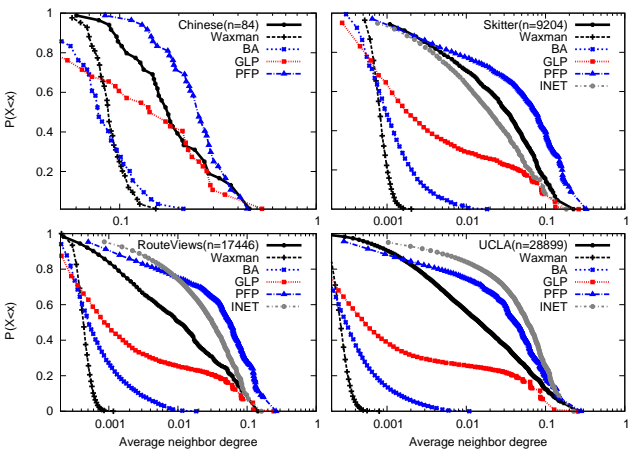

**Figure 3: Comparison of average neighbor connectivity CCDFs.**

Neighbor connectivity has been far less studied than node degree, although it is very important to match local interconnection among a node's neighbors when reproducing the topological structure of the Internet [20]. Figure 3 shows the CCDF of the average neighbor degrees for all topologies. We observe that Waxman, BA and GLP all underestimate the local interconnection structures around nodes due to their way of modeling node interconnections. Note that BA and GLP typically generate graphs with far less links than the observed topologies so they underestimate neighbor degrees on average.

For the larger topologies, i.e. RouteViews and UCLA, PFP and Inet typically overestimate the neighbor connectivity, as they both place a large number of inter-As links in the core. In addition, the shapes of the neighbor connectivity CCDF differ for the larger topologies: Inet and PFP have two regimes, one for highly connected nodes (those with larger neighbor connectivity), and another for low-degree nodes. On the other hand, observed topologies have a smooth region for the high-degree nodes followed by a rather stable region caused by similar degree nodes. We observe that the highest degree nodes in the UCLA topology have very high values of neighbor connectivity. This is consistent with the belief that tier-1 providers are densely meshed. In summary, existing topology generators do not reproduce local interconnection behavior well, even though it is an important aspect of today's AS topology.

### 7.2.3  Clustering coefficients

Like the average neighbor connectivity, the clustering coefficient gives information about local connectivity of the nodes. It is important to reproduce clustering due to its impact on the local robustness in the graph: nodes with higher local clustering have increased local path diversity [20]. Clustering properties of a graph can directly affect simulations on performance of multipath and resilience of overlay routing.

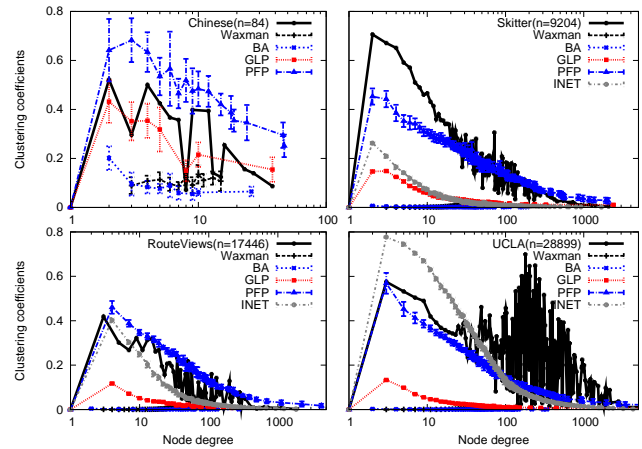

**Figure 4: Comparison of clustering coefficients.**

Figure 4 displays the clustering coefficients of all nodes in the topologies. Error bars indicate 95% confidence intervals around the mean values of the 10 topologies from each generator. We observe that Waxman and BA significantly underestimate clustering, consistent with their simplistic way of connecting nodes. GLP approximates the clustering of the Chinese topology quite well but fails in the case of the larger observed topologies. PFP and Inet capture clustering reasonably well compared to the other topology generators. However, Inet does not reproduce the tail of the distribution well due to the randomness factor in its model for edge addition once the core is fully meshed.

We also observe that for medium degree nodes, clustering coefficients display rather high variability which increases with the size of the observed topologies. This behavior seems to be a property of the observed AS topology of the Internet (Section 8), and not only an artifact of the incompleteness of observed AS topologies.

In summary, all topology generators fail to properly capture clustering, typically underestimating local connectivity. Only Inet for the UCLA topology overestimates connectivity of low-degree nodes while still underestimating it for high-degree nodes. Current topology generators do not seem to have good models of local node connectivity.

### 7.2.4 Rich-club connectivity

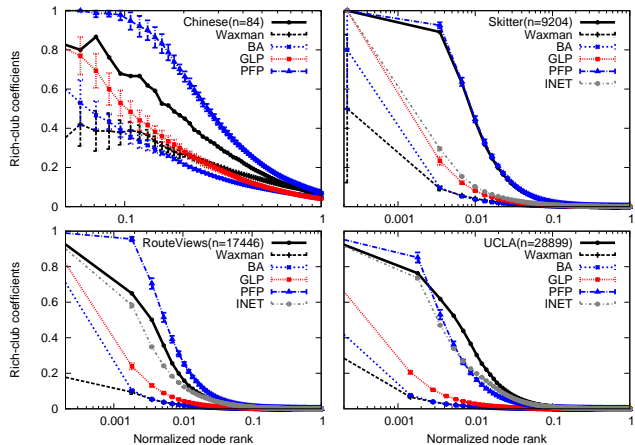

**Figure 5: Comparison of rich-club connectivity coefficients**

Rich-club connectivity gives information about how well-connected nodes of high degree are among themselves. Figure 5 makes it clear that the cores of the observed topologies are very close to a full mesh, with values close to 1 on the left of the graphs. The error bars again indicate the 95% confidence intervals around the mean values of the different instances of the generated topologies. Waxman and BA perform poorly for this metric in general. Only PFP and Inet generate topologies with a dense enough core compared to the observed topologies. Given the emphasis that PFP gives to the rich-club connectivity, it overestimates it in the case of the Chinese and RouteViews topologies. Inet performs well due to its emphasis on a highly connected core, especially for larger topologies where data has been collected across multiple peering points.

In summary, most topology generators underestimate the importance of rich-club connectivity of the AS topology. PFP is the only topology generator that emphasizes the importance of the dense core of the AS topology.

### 7.2.5 Shortest path distributions

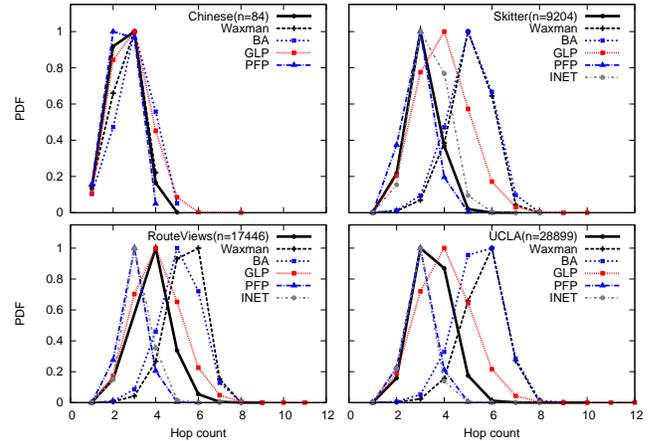

**Figure 6: Comparison of shortest path distributions (number of hops).**

Figure 6 displays the distributions of shortest path length. Apart from BA, most topology generators approximate the shortest path length distribution of the Chinese graph quite well due to its small size. For the other topologies, PFP and Inet generally underestimate the path length distribution while Waxman and BA overestimate. Particular generators seem to capture the path length distribution for particular topologies well: PFP matches that for Skitter well and GLP is close for Routeviews. Inet and PFP both do a better job for UCLA than for RouteViews but both still underestimate the distribution.

In summary, shortest path length is not well captured by any topology generator. Given the poor match of generators to local connectivity metrics, this is not surprising.

### 7.2.6 Spectrum

The spectrum of the normalized Laplacian matrix is a powerful tool for characterizing properties of a graph. If two graphs have the same spectrum, they have the same topological structure.

Figure 7 displays the CDF of the eigenvalues computed from the normalized Laplacian matrix of each topology.

As with other topological metrics, Inet and PFP perform best. The difference between the topology generators is most easily observed around the eigenvalues equal to 1. These eigenvalues play a special role as they

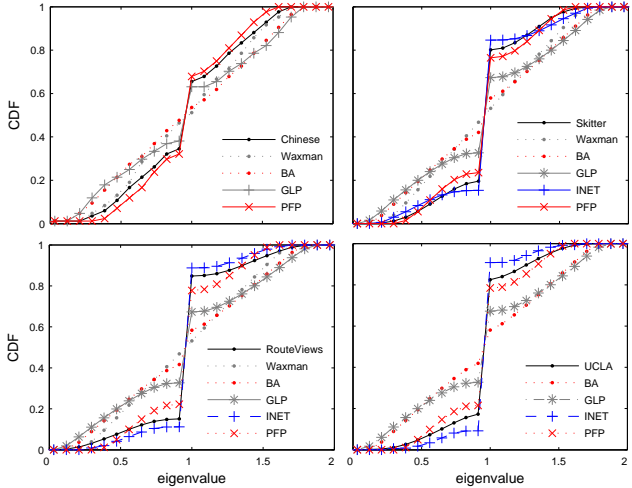

**Figure 7:** Comparison of cumulative distributions of eigenvalues (from normalized laplacian).

indicate repeated duplications of topological patterns within the network. By duplication, we mean different nodes having the same set of neighbors giving their induced subgraphs the same structure. Through repeated duplication, one can create networks with high multiplicity of eigenvalue 1 [2]. Further, if a network is bipartite, i.e., it consists of two connected parts with no links between nodes of the same part, then its spectrum will be symmetric about 1. This phenomenon can also arise through repeated structure duplication

We observe that the spectra have a high degree of symmetry around the eigenvalue 1, and so the observed AS topologies appear close in spectral terms to a bipartite graph. In the AS topology many ASs share a similar set of upstream ASs without being directly connected to each other. Inet and PFP are good examples of topology generators where this strategy is implemented. Note that the simple preferential attachment model of BA does not reproduce the eigenvalues around 1 very well. In the simple BA model, new nodes connect randomly to a given number of existing nodes, favoring connections to high degree nodes. In the Internet in contrast, although small ASs may tend to connect to large upstream providers, they might not connect preferentially to the largest ones, connecting instead to national or regional providers. In summary, these results provide further evidence that the interconnection structure of the AS topology is more complex than current models assume.

### 7.3 Measures of similarity

In Section 7.2, we presented visual evidence for the (dis)similarity both among topology generators and between generators and observed topologies. In this section we present a more objective approach, based on the statistical distance metrics described in Section 6: the Kolmogorov-Smirnov (KS) distance and the Kullback-Leibler (KL) divergence.

In the following tables, the values of the distances and the standard deviations are shown for the topological metrics with distributions: node degree, neighbor connectivity, clustering coefficient, and rich-club coefficient. We provide the average values of the statistical distances and the standard deviation around the average over the 10 topologies generated by each topology generator. When no deviation is shown, it was $< 0.01$.

**Table 2:** Statistical distances for Chinese vs. synthetic topologies.

|        | Node degree    |              | Neighbor connectivity |              |
|--------|----------------|--------------|-----------------------|--------------|
|        | KS distance    | KL divergence| KS distance           | KL divergence|
| Waxman | 0.27±0.07      | **0.6±0.1**  | 0.75±0.03             | 27.4±4.1     |
| BA     | **0.12±0.03**  | 3.5±1.8      | 0.74±0.07             | 18.4±8.1     |
| GLP    | 0.24±0.08      | 0.64±0.31    | **0.41±0.08**         | 1.18±0.72    |
| PFP    | 0.17±0.04      | 1.45±0.48    | 0.51±0.07             | **0.85±0.25**|
|        | Clus. Coefficients |          | Rich-Club Coefficients |             |
|        | KS distance    | KL divergence| KS distance           | KL divergence|
| Waxman | 0.61±0.03      | 22.31±4.5    | 0.22±3.5              | 4.2±2.8      |
| BA     | 0.65±0.1       | 13.5±5.2     | 0.28±0.01             | 2.78±1.4     |
| GLP    | **0.31±0.05**  | 1.08±.6      | 0.26±0.04             | 0.34±0.19    |
| PFP    | 0.32±0.11      | **0.34±0.14**| **0.12±0.01**         | **0.11±0.02**|

Both statistical measures globally confirm the visual inspection of Section 7.2: more recent topology generators produce topologies whose properties are closer to the observed topologies. Table 2 provides the KS and KL results for topology generators against the Chinese topology for the four chosen topological metrics. Topology generators do not show improvement for the node degree. However, for the other three metrics successive topology generators do show improvement. Overall, the PFP and GLP model both have small relative distances to the Chinese dataset, due to the small size of the dataset, the presence of high degree nodes and fewer inter-AS connections.

Table 3 displays the results of the statistical measures for results against the Skitter topology. We observe a particularly good match of the node degree distribution by Inet. PFP outperforms all other topology generators for the clustering coefficients and the rich-club coefficients, consistent with the visual inspection.

Statistical distances for RouteViews (Table 4) show that Inet again better matches the node degree distribution. GLP and Inet both perform better than other generators for neighbor connectivity. PFP performs better than the others on the clustering coefficients. On the other hand, none of the generators manages to obtain a close distance for the rich-club coefficients. In Figure 5,

**Table 3: Statistical distances for Skitter vs. synthetic topologies.**

|        | Node degree | | Neighbor connectivity | |
|--------|-------------|----|----------------------|----|
|        | KS distance | KL divergence | KS distance | KL divergence |
| Waxman | 0.54±0.04 | 2.27±0.15 | 0.99±0.01 | 44.48±0.08 |
| BA     | 0.41±0.02 | 17.1±2.6 | 0.99±0.01 | 44.7±0.25 |
| GLP    | 0.31±0.06 | 17.42±4.1 | 0.31 | 2.16 |
| Inet   | **0.075±0.02** | **4.13** | 0.40±0.02 | **1.82±0.31** |
| PFP    | 0.13±0.03 | 18.2±2.31 | **0.13±0.05** | 18.2±2.21 |
|        | Clust. Coefficients | | Rich-Club Coefficients | |
|        | KS distance | KL divergence | KS distance | KL divergence |
| Waxman | 0.91±0.02 | 40.62±1.2 | 0.2±0.05 | 6.75±1.3 |
| BA     | 0.9±0.05 | 44.62±0.12 | 0.37±0.09 | 7.34±1.21 |
| GLP    | 0.7±0.02 | 19.12±1.8 | 0.3±0.01 | 4.34±.45 |
| INET   | 0.74±0.01 | 11.34±1.23 | 0.25 | 3.82±0.2 |
| PFP    | **0.09±0.02** | **0.59±0.19** | **0.03** | **0.91±0.14** |

**Table 4: Statistical distances for RouteViews vs. synthetic topologies.**

|        | Node degree | | Neighbor connectivity | |
|--------|-------------|----|----------------------|----|
|        | KS distance | KL divergence | KS distance | KL divergence |
| Waxman | 0.5±0.03 | 50.77±0.01 | 0.94±0.01 | 42.68±0.25 |
| BA     | 0.2±0.02 | 50.74±0.01 | 0.94±0.01 | 42.91±0.34 |
| GLP    | 0.18±0.03 | 50.73±0.01 | **0.12±0.02** | **0.1±0.02** |
| Inet   | **0.07** | **9.92** | 0.23±0.02 | 0.2±0.02 |
| PFP    | 0.11±0.03 | 50.7 | 0.62±0.02 | 1.25±0.07 |
|        | Clust. Coefficients | | Rich-Club Coefficients | |
|        | KS distance | KL divergence | KS distance | KL divergence |
| Waxman | 0.83±0.05 | 39.4±1.2 | 0.97 | 42.23±0.43 |
| BA     | 0.96±0.01 | 44.08±0.21 | 0.97 | 43.07±0.6 |
| GLP    | 0.58±0.02 | 12.9±0.65 | 0.96 | 40.7±0.9 |
| INET   | 0.39±0.01 | 1.35±0.2 | 0.93 | 34.18±1.1 |
| PFP    | **0.32±0.06** | **0.21±0.03** | **0.92** | **27.4±2.45** |

**Table 5: Statistical distances for UCLA vs. synthetic topologies.**

|        | Node degree | | Neighbor connectivity | |
|--------|-------------|----|----------------------|----|
|        | KS distance | KL divergence | KS distance | KL divergence |
| Waxman | 0.52±0.01 | **1.33±0.9** | 0.99±0.01 | 46.31±1.3 |
| BA     | 0.17±0.03 | 2.15±0.8 | 0.99±0.01 | 46.42±0.7 |
| GLP    | 0.18±0.05 | 2.21±0.7 | 0.32±0.03 | 0.63±0.04 |
| Inet   | 0.2±0.02 | 5.34 | **0.29±0.01** | **0.41±0.01** |
| PFP    | **0.12±0.03** | 2.17±0.8 | 0.48±0.05 | 0.83±0.21 |
|        | Clust. Coefficients | | Rich-Club Coefficients | |
|        | KS distance | KL divergence | KS distance | KL divergence |
| Waxman | 0.93±0.02 | 44.2±0.34 | 0.31 | 14.5±4.32 |
| BA     | 0.99±0.01 | 45.42 | 0.5 | 14.32±2.3 |
| GLP    | 0.82±0.01 | 33.32±0.9 | 0.42±0.01 | 8.9±1.2 |
| INET   | **0.38±0.01** | **0.53±0.01** | **0.13** | **2.85±0.12** |
| PFP    | 0.38±0.02 | 0.79±0.15 | 0.16 | 3.23±0.4 |

Inet seemed to be close to RouteViews for rich-club coefficients, but this is not supported by the statistical distances. The behavior for rich-club connectivity is surprising, especially for PFP which is highly biased towards reproducing rich-club connectivity. We believe this is due mainly to the addition of many extra peering links in this dataset, which was not captured by model designers.

Statistical distance results for UCLA (Table 5) reveal a more complex picture. For node degrees, no generator seems to outperform the others, although Inet does perform worst. GLP, Inet and PFP perform equally well on the neighbor connectivity. For clustering coefficients and rich-club connectivity, Inet and PFP perform better than the others.

Visual inspection of Section 7.2 seemed to suggest that each successive topology generator introduced improvements in their matching of observed AS topologies. Waxman and BA perform poorly both in visual inspection and in the statistical distances. The KL divergences clarify the difference of the two distributions across all the values and hence minimize the effects of local differences at certain values.

Our statistical measures show that apparent visual closeness of two distributions does not mean close distance in distributional terms, due partly to the use of logarithmic scale axes. Improvements in successive topology generators are not consistent across all metrics and across all observed topologies. Nonetheless, most of the time the most recent generators, Inet and PFP, do outperform the other topology generators. This indicates that more attention should be given on capturing the effects of peering links in the core and at the edge of the AS topology, as this is the significant difference between these two generators and the older Waxman and BA generators.

## 8. MULTIPLE VANTAGE POINTS

The previous section studied in detail *how well* topology generators capture the properties of observed AS topologies. In this section, we will argue about *why* topology generators capture different propeties of observed AS topologies with varying degrees of success. To that end we examine the impact on the metrics of the number of vantage points from which BGP data is collected. For our analysis we collected BGP data from over 40 RouteViews peering points, for a period of 6 months from May 2007. This time period was chosen to be the same as that used to build the UCLA dataset.

Zhang *et al.* [36] also examine the impact of the selection of route monitors on topology visibility and the consequences on AS relationship inference and AS-level path prediction. They analyze a range of monitor selection schemes and their influence on the number of observed links as well as network properties. They suggest that the accuracy of AS relationship inference may decrease as the number of monitors increases, and go on to quantify the improvements in identifying AS relationships and anomaly detection in the data. In contrast, our work focuses on understanding the underlying effect placement of vantage points has on inferring both the network topology and its associated dynamics. We are also interested in examining the distortion of local topological properties by using a different number of vantage points.

Table 6 shows the values of the topological metrics the same way as in Table 1, for AS topologies obtained from different numbers of observation points. When comparing the AS topologies using 1 and 10 observation points, we see a significant increase in the number of nodes and links. Hence, one might also expect a significant difference in the other metrics, and indeed, the maximum node degree almost triples and the number of fully-meshed nodes almost doubles. As a consequence, the size of the core increases, indicated by the maximum coreness value. In turn, the number of shortest paths crossing the core also increases as indicated by the maximum betweenness. On the other hand, we see that going from 1 to 10 observation points slightly decreases the value of the clustering coefficient. Most probably this is because with 10 observation points we discover more of the core than the edge of the network, which does not contribute to increase the overall value of the clustering coefficient. With 25 or more observation points the links on the edge of the network are also discovered, contributing to the increase of the value of the clustering coefficient. This behavior is confirmed by a slight decrease of the value of the maximum betweenness from 10 to 25 observation points.

Preferential attachment models originate in the belief that small ASs tend to connect to large upstream ASs, leading to a disassortative network. Although the value of the assortativity coefficient is negative for the AS topology, it is not affected by an increase in the number of observation points. The links added by increasing the number of observation points seem to be neutral for the assortativity of the AS topology. One implication is that the links that can be discovered by using more observation points do not preferentially interconnect ASs of any particular degree. We conjecture that this is due to the type of peering relationships that are missed. If node degrees give an indication of the likely type of peering relationship, then we suggest that BGP does not preferentially miss peer-peer relationships, which are believed to be more difficult to observe that customer-provider ones [6].

We now turn in more detail to the effect of the number of peering points on four particular topological metrics (see Figure 8). The addition of observation points mostly affects node degree distribution for high degree nodes. As we increase the number of observation points, we see that on average the neighbors of a node will have a higher degree. However, this does not hold for nodes whose neighbors already have high degrees (left part of Figure 8. Those nodes correspond to stub networks connected to very well interconnected upstream providers. For the clustering coefficient, when moving from one to several observation points, the difference is striking. For all node degrees, the clustering coefficient significantly increases. On the other hand, when moving from a few peerings to many, the difference appears most for high degree nodes. This illustrates the better observability of links in the core compared to the edge of the network. Rich-club connectivity confirms the previous observations in that adding a few observation points is enough to discover the core links.

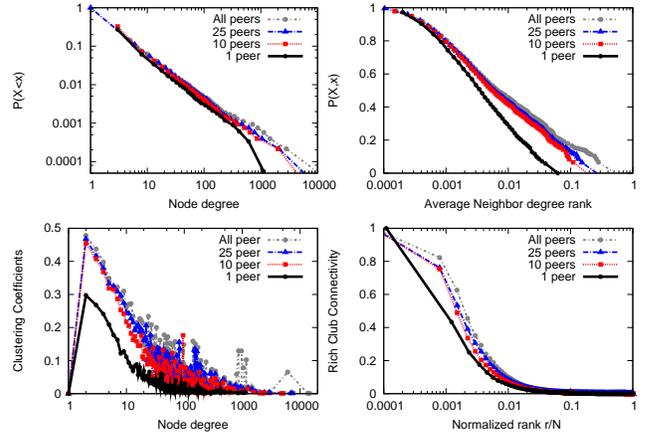

Figure 8: Comparison of effects of the number of peering points.

In this section we have illustrated the importance of relying on a sufficiently large number of observation points in order to properly capture the actual properties of the AS topology. Using only a few observation points has led researchers to simplify the complexity of the interconnection structure between ASs. The improper AS topology on which researchers have relied has caused the creation of topology generators that underestimate this interconnection structure between ASs. Our results show that researchers must use rich datasets for an accurate understanding of the Internet AS topology.

## 9. CONCLUSIONS

In this paper we evaluated the existing AS topology generation models, by comparing them with several observed AS topologies. For this evaluation, we relied on a wide set of topological metrics and statistical measures to carry our comparison as objectively as possible. Our analysis revealed that:

- Increasing the number of observation points causes deviation from strict degree power-law scaling. Existing topology generation models overemphasize the preferential attachment mechanism and the resulting node degree distribution. The power-law assumption is thus an artefact of incomplete datasets, rather than a property of the AS-level topology.

Table 6: Comparison of AS topology datasets from multiple peering points.

| Topology | Nodes | Links | Avg. deg. | Max. degree | Top clique size | Max. betweenness | Max. coreness | Assort. coef. | Clust. coef. | Max. closeness |
|---|---|---|---|---|---|---|---|---|---|---|
| 1 peer | 17,952 | 34,617 | 3.86 | 980 | 4 | 35,069,182 | 9 | -0.18 | 0.008 | <0.01 |
| 10 peers | 27,838 | 64,717 | 4.65 | 2,731 | 7 | 52,862,315 | 20 | -0.18 | 0.007 | <0.01 |
| 25 peers | 27,885 | 67,659 | 4.85 | 2,808 | 7 | 49,798,002 | 25 | -0.19 | 0.01 | <0.01 |
| All peers | 27,924 | 70,064 | 5.02 | 3,371 | 7 | 70,142,726 | 30 | -0.18 | 0.01 | <0.01 |

- In addition to clustering and centrality properties, the highly meshed core of the Internet AS topology must be considered in order to generate representative synthetic topologies.

- The successive improvements in topology generation models seems to result from improved available datasets. Knowing that incomplete datasets were the cause for simplistic topology generation models, we expect that the new generation of topology models will take into account the insights gained in this paper.

Our findings indicate that future work in this area should consider the geographical extent of the AS graphs, the AS sizes, multiple peerings between ISPs, policy routing and topology dynamics. Future AS topology generators should also permit the addition of metadata such as peering relationships and relative importance of nodes.

## Acknowledgments

We wish to acknowledge Richard Gibbens and Andrew Thomason for giving insight into the topology models and comparison techniques. We would also like to thank Jon Crowcroft, Tim Griffin, Vivek Mhatre, and Tao Ye for their feedback on earlier versions of this work. Priya Mahadevan provided some of the code for calculation of connectivity metrics and Belinda Chiera helped in preparing BGP data. This work is conducted as part of the EPSRC UKLIGHT/MASTS project under grants GR/T10503/01 and GR/T10510/03.